\documentclass[12pt]{article}

  \usepackage{graphicx}
  \usepackage{epsfig}
  \usepackage{amssymb}
  \usepackage{subfigure}
  \usepackage{axodraw}
\evensidemargin - .5truecm
\begin{document}

\newcommand{\be}{\begin{equation}}
\newcommand{\ee}{\end{equation}}
\newcommand{\nn}{\nonumber}
\newcommand{\bea}{\begin{eqnarray}}
\newcommand{\eea}{\end{eqnarray}}
\newcommand{\bfig}{\begin{figure}}
\newcommand{\efig}{\end{figure}}
\newcommand{\bc}{\begin{center}}
\newcommand{\ec}{\end{center}}
\def\ad{\dot{\alpha}}
\def\ov{\overline}
\def\hlf{\frac{1}{2}}
\def\qrt{\frac{1}{4}}
\def\as{\alpha_s}
\def\at{\alpha_t}
\def\ab{\alpha_b}
\def\sq2{\sqrt{2}}
\newcommand{\smallz}{{\scriptscriptstyle Z}} %
\newcommand{\mz}{m_\smallz}
\newcommand{\smallw}{{\scriptscriptstyle W}}
\newcommand{\mw}{m_\smallw} 
\newcommand{\smallh}{{\scriptscriptstyle H}}
\newcommand{\mh}{m_\smallh}
\newcommand{\mt}{m_t}
\newcommand{\wh}{w_\smallh}
\def\th{t_\smallh}
\def\zh{z_\smallh}
\newcommand{\Mvariable}[1]{#1}
\newcommand{\Mfunction}[1]{#1}
\newcommand{\Muserfunction}[1]{#1}

\newenvironment{appendletterA}
 {
  \typeout{ Starting Appendix \thesection }
  \setcounter{section}{0}
  \setcounter{equation}{0}
  \renewcommand{\theequation}{A\arabic{equation}}
 }{
  \typeout{Appendix done}
 }
\newenvironment{appendletterB}
 {
  \typeout{ Starting Appendix \thesection }
  \setcounter{equation}{0}
  \renewcommand{\theequation}{B\arabic{equation}}
 }{
  \typeout{Appendix done}
 }

%
%
%
%
%

\begin{titlepage}
\nopagebreak
{\flushright{
        \begin{minipage}{5cm}
         Rome1-1442/06 \\
         IFIC/06-31 \\
         RM3-TH/06-24 \\
         IFUM-880-FT \\ 
        {\tt hep-ph/0611266}
        \end{minipage}        }

}
\renewcommand{\thefootnote}{\fnsymbol{footnote}}
\vskip 1.cm
\begin{center}
\boldmath
{\Large\bf Analytic Results for Virtual QCD Corrections \\[7pt]
to Higgs Production and Decay}\unboldmath
\vskip 1.cm
{\large  U.~Aglietti\footnote{Email: Ugo.Aglietti@roma1.infn.it}},
\vskip .2cm
{\it Dipartimento di Fisica, Universit\`a di Roma ``La Sapienza'' and
INFN, Sezione di Roma, P.le Aldo Moro~2, I-00185 Rome, Italy} 
\vskip .2cm
{\large  R. Bonciani\footnote{Email:
Roberto.Bonciani@ific.uv.es}},
\vskip .2cm
{\it Departament de F\'{\i}sica Te\`orica, 
IFIC, CSIC -- Universitat de 
Val\`encia, \\
E-46071 Val\`encia, Spain}
\vskip .2cm
{\large G.~Degrassi\footnote{Email: degrassi@fis.uniroma3.it}},
\vskip .2cm
{\it Dipartimento di Fisica, Universit\`a di Roma Tre and 
INFN, Sezione di Roma Tre, \\ Via della Vasca Navale~84, I-00146 Rome, Italy} 
\vskip .2cm
{\large A.~Vicini\footnote{Email: Alessandro.Vicini@mi.infn.it}}
\vskip .2cm
{\it Dipartimento di Fisica, Universit\`a di Milano and
INFN, Sezione di Milano, \\
Via Celoria 16, I--20133 Milano, Italy} 
\end{center}
\vskip 0.7cm

\begin{abstract}
We consider the production of a Higgs boson via gluon-fusion 
and its  decay into two photons.
We compute  the NLO virtual QCD corrections to these processes in a general
framework in which the coupling of the Higgs boson to the external particles 
is mediated  by a colored fermion and a colored scalar. We present compact 
analytic results for these two-loop corrections that are expressed in terms 
of  Harmonic Polylogarithms.  The expansion of these corrections 
in the low and high Higgs mass regimes, as well as the expression of the new
Master Integrals which appear in the reduction of the two-loop
amplitudes, are also provided. For the fermionic contribution,
we provide an independent check of the results already present in the 
literature concerning the Higgs boson and the production and decay of
a pseudoscalar particle. 


\end{abstract}
\vfill
\end{titlepage}    
\setcounter{footnote}{0}


\section{Introduction}

The Higgs searches program at the TEVATRON and at the LHC
requires from the theoretical side
the highest possible level of accuracy in the prediction of the
production cross-sections and of all the decay channels. Over the years
a lot of effort has been devoted to the study of the QCD, and also EW,
corrections to the various production mechanisms and decays in the
Standard Model and beyond (for a recent review see Ref.\cite{Dj}).

The gluon-fusion process $gg\to H+X$ \cite{H2gQCD0}
is the dominant production mechanism. Its present knowledge includes
the NLO \cite{H2gQCD1, QCDg2,HK} and NNLO QCD corrections \cite{H2gQCD2} 
and  the two-loop EW corrections \cite{DjG,ABDV,DM1}.
The QCD corrections to Higgs production at finite transverse momentum
have also been discussed \cite{QTnlo}.
While the NLO QCD corrections and the two-loop EW light fermion contribution 
are known completely, namely for arbitrary value of the Higgs mass and of the
other relevant particles in the loops, the NNLO QCD corrections are only
known in the heavy top limit while the result for the two-loop EW top 
contribution is valid only for intermediate Higgs mass, i.e. $\mh \le 2 \mw$.

The Higgs decay $H\to\gamma\gamma$ \cite{oneloop} is, for light values of the 
boson mass, a very promising channel. It has been studied in great detail
including the NLO QCD \cite{QCD2loop,ftt} and the two-loop EW corrections
\cite{EW2lmt, ABDV, DM2,FKS}. The NLO QCD corrections are now known in a closed
analytic form \cite{ftt,HK}, while for the EW corrections their knowledge
is similar to that of the gluon fusion process.

Given the importance of the Higgs physics program, it is highly desirable to 
have the radiative corrections to the various  reactions expressed in 
analytic form that can be easily implemented in computer codes. With respect
to this, it should be recalled that the complete result concerning the NLO QCD
corrections to the gluon fusion process has been reported in Ref.\cite{QCDg2}
via a rather lengthy formula expressed in terms of  a one-dimensional integral
representation. Actually the calculation of the two-loop light-fermion
EW corrections to the Higgs production and decay \cite{ABDV} has shown that 
corrections of this kind can be  calculated analytically, expressing the 
results in terms of Harmonic Polylogarithms (HPL)~\cite{HPLs},
a generalization of Nielsen's polylogarithms, and an extension of the
HPL, the so-called Generalized Harmonic Polylogarithms (GHPL)~\cite{AB2}.
The idea lying behind the introduction of (G)HPLs is to express a given
integral coming from the calculation of a Feynman diagram in a unique
and non-redundant way as a linear combination of a minimal set of
independent transcendental functions. These functions are expressed as
repeated integrations over a starting set of basis functions and this
set depends strongly on the problem one has to solve, being connected
directly to the threshold structure of the diagrams under consideration.

An inspection of the threshold structure of the NLO QCD corrections to
the gluon-fusion process and to $H\to\gamma\gamma$ decay shows that
these corrections can be fully expressed in terms of the original set of 
HPLs  introduced in \cite{HPLs}.
A  FORTRAN program \cite{GR} and a Mathematica package \cite{Mai} 
that efficiently evaluate  these functions are available.

The aim of this paper is to provide  analytic expressions, in terms of HPLs,
for the NLO  QCD corrections to the Higgs  production cross 
section via gluon fusion, i.e. $gg\to H$, in a general form that
can be applied both to the SM and to models beyond it, and, moreover, to 
provide an independent check for the formulas already present in the 
literature.   The production mechanism is assumed to be mediated by 
colored fermion and scalar 
particles. As a byproduct we also present the NLO QCD corrections to the 
Higgs  decay into two photons, i.e. $H \to\gamma\gamma$. A similar
project has been carried out in Ref.\cite{HK}. There, the authors started
from the result of Ref.\cite{QCDg2}\footnote{In Ref.\cite{QCDg2} the QCD
corrections were considered only for the fermion contribution.}
expressed as a one-dimensional integral representation. Expanding this result
in a power series, employing the theorem that two analytic functions are the 
same if their Taylor series are the same, they were able to rewrite it  
in terms of HPLs. In our case we explicitly compute all the relevant 
Feynman diagrams, expressing the result in term 
of HPLs. The calculational techniques we employed are the Laporta
algorithm \cite{Lap} for the reduction to Master Integrals (MIs) and the 
differential equation method \cite{DiffEq} for their calculation 
(the calculation is implemented in FORM \cite{FORM} codes).
To complete an independent check of the results presented in Ref.\cite{QCDg2}
we also computed the NLO QCD corrections to the pseudoscalar production
and decay, i.e. $gg\to A,\: A \to \gamma\gamma$.

The paper is organized as follows: in Section \ref{sec2} we discuss the QCD
corrections to the decay width $H \to \gamma\gamma$. Section \ref{sec3} 
is devoted to the study of the Higgs production via the gluon fusion
mechanism. The following section contains the analytic  expressions
for the virtual QCD corrections to the fermionic contribution in 
$gg\to A,\: A \to \gamma\gamma$. Finally we present our conclusions.
We include also two Appendices. In the first one we collect the expansions of
the relevant functions in the two regimes: for Higgs mass much lighter than
the particles mediating the Higgs interaction with the vector bosons and in 
the  opposite case.
In the second Appendix we collect the MIs not already present
in the literature, that enters the calculation of the NLO QCD corrections.


\section{The $H\to\gamma\gamma$ Decay Width \label{sec2}}

We begin by considering the decay width $H\to\gamma\gamma$. Being the
Higgs boson electrically neutral its coupling to the photon is mediated at 
the loop-level by charged particles. For the latters we assume a vector
boson neutral under $SU(N_c)$, a fermion and a scalar particle  
in a generic $R_{1/2},\, R_0$ $SU(N_c)$ representation, respectively, 
whose coupling's strengths to the Higgs are: 
\be
HVV = g \,\lambda_1 \,\mw, ~~~~~~~~~~~
HFF= g \,\lambda_{1/2}\, \frac{m_{1/2}}{2\,\mw},
~~~~~~~~~~~HSS = g \,\lambda_{0}\, \frac{A^2}{\mw}~,
\ee
where $g$ is the $SU(2)$ coupling, $\mw$ is the W mass, $m_{1/2}$ is the
fermion mass, $A$ is a generic coupling 
with the dimension of mass and 
$\lambda_i$ are numerical coefficients\footnote{The SM is recovered with 
$\lambda_1 = \lambda_{1/2}=1, \: \lambda_0 = 0$, $N_c=3$ and 
$R_{1/2} = {\mathbf 3}$.}. 

The partial decay width for the reaction $H\to\gamma\gamma$ can be written as:
\be
\Gamma(H \to \gamma \gamma) = \frac{G_{\mu} \alpha^2 m_H^3}{128 \sqrt{2}
\pi^3} \left| {\mathcal F} \right| ^2 ~,
\label{width}
\ee
where the  function $ {\mathcal F}$ can be organized with respect to the lowest
order term and its QCD corrections as:
\be
{\mathcal F} = \lambda_1 \, Q_1^2 \, N_1\,{\mathcal F}_{1} +    
        \lambda_{1/2} \,  Q_{1/2}^2 \,N_{1/2}    {\mathcal F}_{1/2} +
 \lambda_{0} \,  Q_{0}^2 \, N_0  \frac{A^2}{m_0^2}\,  {\mathcal F}_{0} \, ,
\ee
where  $m_0$ is the mass of the scalar particle, while $Q_i$ and $N_i$, 
$i=0, 1/2,1$, are the electric charges and  the 
representation numbers under $SU(N_c)$ of the scalar, fermion and vector boson 
particles, respectively.

Writing:
\be
{\mathcal F}_i = {\mathcal F}^{(1l)}_i + {\mathcal F}^{(2l)}_i + \dots
\ee
we have at the one-loop level
\bea
{\mathcal F}^{(1l)}_{1} & = & 2 ( 1 + 6 y_1) - 12 y_1 (1 - 2 y_1)
                                 \, H (0,0,x_1) \, , \label{eq:2} \\
{\mathcal F}^{(1l)}_{1/2} & = & - 4 y_{1/2} 
 \left[ 2 - \left( 1 -4 y_{1/2} \right)  \, H(0,0,x_{1/2}) \right] \, , 
\label{eq:3} \\
{\mathcal F}^{(1l)}_{0} & = & 4 y_0 \left[ 1 + 2 \, y_0\,
                 H(0,0,x_{0}) \right]~.
\label{eq:4}
\eea
In Eqs.({\ref{eq:2}-\ref{eq:4}) 
\be
y_i \equiv \frac{m_i^2}{\mh^2}, ~~~~~~~~
x_i \equiv \frac{\sqrt{1- 4 y_i} - 1}{\sqrt{1- 4 y_i} + 1}  \, ,
\label{defx}
\ee
with  $m_1$ the mass of the vector particle and, employing the
standard notation for the HPLs,  $H(0,0,z)$ labels a HPL of weight 2
that results to be\footnote{All the analytic  continuations are obtained with 
the replacement $-\mh^2 \to -\mh^2 - i \epsilon$}
\be 
H (0,0,z ) = \frac{1}{2} \log^2 (z)~.
\ee

The QCD corrections to the lowest order result can be written as
\be
{\mathcal F}_{QCD}^{(2l)}  =  \frac{\as}{\pi}  \sum_{i=(0,1/2)} C(R_i)\,  
                      {\mathcal F}_{i}^{(2l)} \, ,
\label{eq:9}
\ee
where $C(R)$ is the Casimir factor of the $R_i$ representation (in particular,
for the fundamental and the adjoint representations of $SU(N_c)$ we have
$C_F = (N_c^2-1)/(2N_c)$ and $C_A = N_c$, respectively).
We consider first the fermion contribution (the relevant Feynman diagrams are 
shown in Fig.~\ref{Hgaga} (a)--(d)).

\begin{figure}
\bc
\epsfig{file=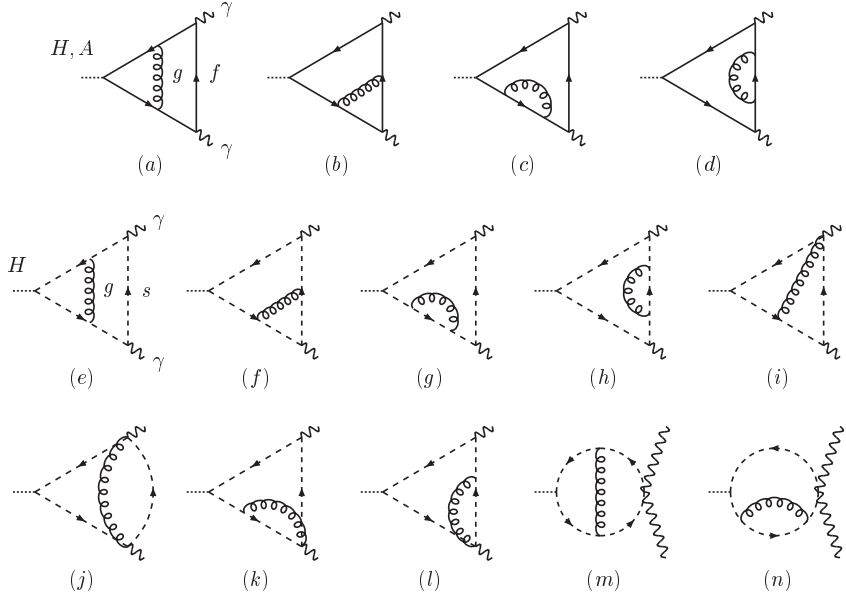,height=8.2cm,width=4.5cm,
        bbllx=220pt,bblly=480pt,bburx=375pt,bbury=763pt}
\caption{\label{Hgaga} \it The Feynman diagrams for the decay process 
$H,A \to \gamma \gamma$. Diagrams (a)--(d) have a fermion, labeled by ``f'', 
running in the loop, while diagrams (e)--(n) have a scalar, labeled by ``s''.}
\ec
\end{figure}

The expression for ${\mathcal F}_{1/2}^{(2l)}$ depends upon the renormalized
mass parameter employed. In the case of $\overline{\mbox{MS}}$ 
quark  masses we have
\be
 {\mathcal F}_{1/2}^{(2l)} =  {\mathcal F}_{1/2}^{(2l,a)}(x_{1/2}) 
                   + {\mathcal F}_{1/2}^{(2l,b)}(x_{1/2}) 
                   \ln \left({m_{1/2}^2 \over \mu^2} \right) \, ,
\ee
where $\mu$ is the 't~Hooft mass and 
\bea
{\mathcal F}_{1/2}^{(2l,a)} (x) & = & 
        \frac{36x}{(x-1)^2} 
      - \frac{4x \left( 1-14x+x^2 \right)}{(x-1)^4} \, \zeta_3 
      - \frac{4x (1+x)}{(x-1)^3} \, H(0,x) \nn\\
& &  
      - \frac{8x \left( 1+9x+x^2 \right)}{(x-1)^4} \, H(0,0,x) 
      + \frac{2x \left( 3+25x-7x^2+3x^3 \right)}{(x-1)^5}
             \,H(0,0,0,x) \nn \\
& &  
      + \frac{4x \left( 1+2x+x^2 \right)}{(x-1)^4} \, 
        \bigl[ \zeta_2 H(0,x) + 4 \, H(0, -1,0,x) - H(0,1,0,x) \bigr] \nn \\
& &   
      + \frac{4x \left( 5-6x+5x^2 \right)}{(x-1)^4} \, H(1,0,0,x)
      - \frac{8x \left( 1+x+x^2+x^3 \right)}{(x-1)^5} \, {\mathcal H}_1(x)
     \, ,
\label{2lQCD}\\
{\mathcal F}_{1/2}^{(2l,b)} (x) & = &
      - \frac{12 x}{(x-1)^2} 
      - \frac{6 x (1+x) }{(x-1)^3} \, H(0,x) 
      + \frac{6 x \left( 1+6x+x^2 \right)}{(x-1)^4} \, H(0,0,x) \, ,
\label{2lQCDpole}
\eea
with
\bea
{\mathcal H}_1(x)  &=&  
        \frac{9}{10}{\zeta_2}^2 + 2 \zeta_3 H(0,x) 
      + \zeta_2 H(0,0,x) + \frac{1}{4}\,H(0,0,0,0,x) 
      + \frac{7}{2}\,H(0,1,0,0,x) \nn \\
& & 
      - 2 \, H(0,-1,0,0,x)  +  4 \, H(0,0,-1,0,x) - H(0,0,1,0,x) \, .
\label{Hfun}
\eea
In Eqs.(\ref{2lQCD},\ref{Hfun})  $\zeta_n \equiv \zeta(n)$ are the Riemann's 
zeta functions.

The expression for ${\mathcal F}_{1/2}^{(2l)}$ in case the one-loop result
is given in terms of on-shell fermion masses is given instead by:
\be
 {\mathcal F}_{1/2}^{(2l)} =  {\mathcal F}_{1/2}^{(2l,a)} (x_{1/2})
                   + \frac43 {\mathcal F}_{1/2}^{(2l,b)}(x_{1/2})~. 
\label{2lQCDos}
\ee
Eq.(\ref{2lQCDos}) is in agreement with the results
presented in \cite{ftt,HK}.

We now present the scalar contribution, ${\mathcal F}_{0}^{(2l)}$, assuming 
that both the mass of the scalar, $m_0$, and the coupling $A$ are 
renormalized in 
the $\overline{\mbox{MS}}$ scheme (the relevant Feynman diagrams are 
shown in Fig.~\ref{Hgaga} (e)--(n)). We find
\be
 {\mathcal F}_{0}^{(2l)} =  {\mathcal F}_{0}^{(2l,a)}(x_{0}) 
                   + \left( {\mathcal F}_{0}^{(2l,b)}(x_{0}) +
                   {\mathcal F}_{0}^{(2l,c)}(x_{0}) \right)
                   \ln \left({m_{0}^2 \over \mu^2} \right) \, ,
\ee
where 
\bea
{\mathcal F}_{0}^{(2l,a)} (x) & = & -\frac{14 x}{(x-1)^2} 
       - \frac{24x^2}{(x-1)^4} \,\zeta_3 
       + \frac{x \left( 3-8x+3x^2 \right) }{(x-1)^3(x+1)} \, H(0,x)
       + \frac{34x^2}{{\left( x-1 \right) }^4} \, H(0,0,x) \nn \\
& &  
       - \frac{8x^2}{(x-1)^4} \, \bigl[ 
              \zeta_2 H(0,x) 
	    + 4 H(0, -1,0,x) 
	    - H(0,1,0,x) 
	    + H(1,0,0,x) \bigr] \nn \\
& &  
       - \frac{2x^2 (5-11x)}{(x-1)^5} \, H(0,0,0,x) 
       + \frac{16x^2 \left( 1+x^2 \right)}{(x-1)^5(x+1)} \, {\mathcal H}_1(x) \, ,
\label{2lsca} \\
{\mathcal F}_{0}^{(2l,b)} (x) & = & 
         \frac{6x^2}{(x-1)^3(x+1)} \, H(0,x)
       - \frac{6x^2}{(x-1)^4} \, H(0,0,x) \, , \\
{\mathcal F}_{0}^{(2l,c)} (x) & = & -\frac34 \,{\mathcal F}_{0}^{(1l)} \, .
\eea

We provide also $ {\mathcal F}_{0}^{(2l)}$ assuming that the mass of the 
scalar is
renormalized on-shell while the $A$ coupling is still given as an
$\overline{\mbox{MS}}$ one. It reads
\be
 {\mathcal F}_{0}^{(2l)} =  {\mathcal F}_{0}^{(2l,a)}(x_{0}) 
                   + \frac{7}{3} \, {\mathcal F}_{0}^{(2l,b)}(x_{0})+ 
{\mathcal F}_{0}^{(2l,c)}(x_{0}) \ln \left({m_{0}^2 \over \mu^2} 
		     \right)~.
\label{2lscaos}
\ee


\section{Virtual Corrections to  $gg \to H$ Production Mechanism \label{sec3}}

In this section we present the analytic expressions for the virtual
two-loop QCD corrections for  Higgs boson production via the gluon fusion
mechanism. 
Being the Higgs boson neutral under $SU(N_c)$, its coupling to the gluons is
mediated by a loop of colored particles. As in Section \ref{sec2}, we consider
a fermion and a scalar particle, that run in the loops.
The Feynman diagrams relevant for the NLO corrections to the production cross
section are shown in Fig.~\ref{ggH}.

\begin{figure}[t]
\bc
\epsfig{file=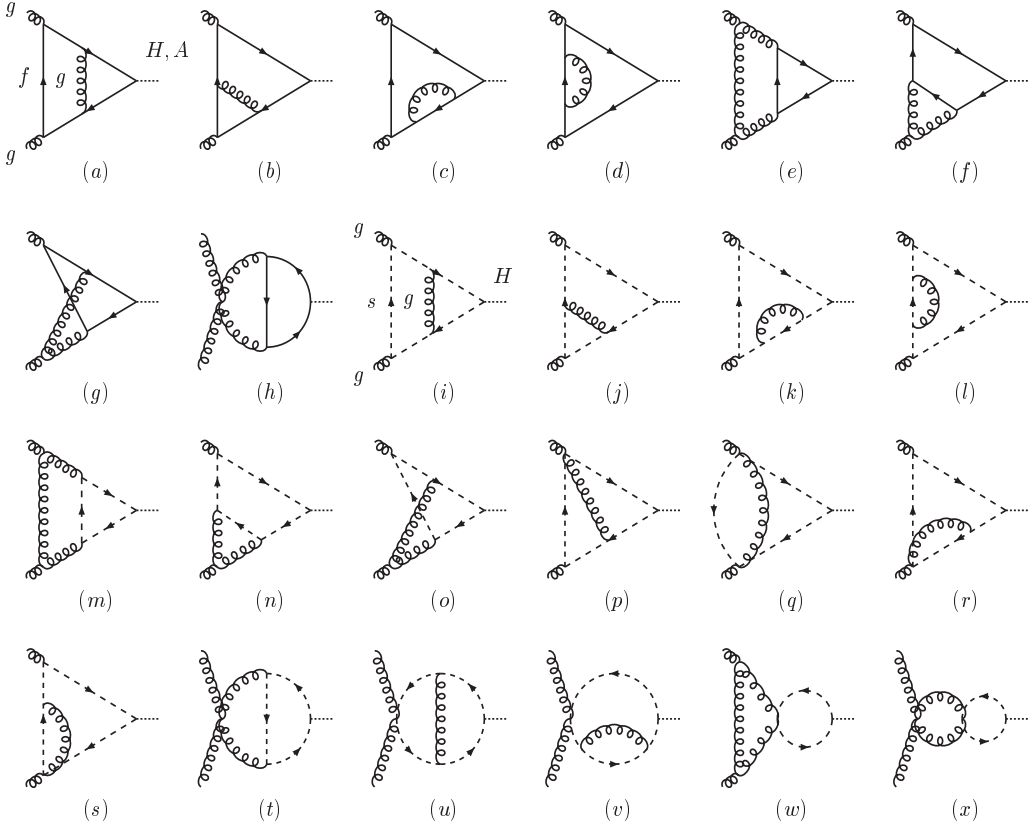,height=11.2cm,width=4.5cm,
        bbllx=220pt,bblly=390pt,bburx=375pt,bbury=763pt}
\caption{\label{ggH} \it The Feynman diagrams for the production mechanism
$gg \to H,A$. Diagrams (a)--(h) have a fermion, labeled by ``f'', 
running in the loop, while diagrams (i)--(x) have a scalar, labeled by ``s''.}
\ec
\end{figure}

The hadronic cross section can be written as:
\bea
\sigma(h_1 + h_2 \to H+X) & = & 
          \sum_{a,b}\int_0^1 dx_1 dx_2 \,\,f_{a,h_1}(x_1,\mu_F^2)\,
         f_{b,h_2}(x_2,\mu_F^2) \times \nonumber\\
& & \times
\int_0^1 dz~ \delta \left(z-\frac{\tau_H}{x_1 x_2} \right)
\hat\sigma_{ab}(z) \, ,
\label{sigmafull}
\eea
where $\tau_H= \mh^2/s$, $\mu_F$ is the factorization scale,
$f_{a,h_i}(x,\mu_F^2)$, the parton density of the colliding hadron $h_i$
for the parton of type $a, \,(a = g,q,\bar{q})$ and $\hat\sigma_{ab}$ the 
cross section for the partonic subprocess $ ab \to H +X$ at the center-of-mass 
energy  $\hat{s}=x_1 x_2 s=\mh^2/z$. The latter can be written as:
\be
\hat\sigma_{ab}(z)=
\sigma^{(0)}\,z \, G_{ab}(z) \, ,
\label{Geq}
\ee
where 
\be
\sigma^{(0)}  =  
\frac{G_\mu \alpha_s^2 (\mu_R^2)}{128\, \sqrt{2} \, \pi}
\left|   \sum_{i=0,1/2} \lambda_i \left(\frac{A^2}{m_0^2} \right)^{1-2 i}\,
T(R_i)\, {\mathcal G}^{(1l)}_{i}
       \right|^2 
\label{ggh}
\ee
is the Born-level contribution 
with ${\mathcal G}^{(1l)}_{i}= {\mathcal F}^{(1l)}_{i} $
and $T(R_i)$ are the matrix normalization factors of the $R_i$ representation
($T(R) =1/2$ for the fundamental  representation of $SU(N_c)$, 
$T(R) =N_c$ for the adjoint one). 

Up to NLO contributions we can write
\be
G_{a b}(z)  =  G_{a b}^{(0)}(z) 
          + \frac{\alpha_s (\mu^2_R)}{\pi} \, G_{a,b}^{(1)}(z) \, ,
\ee
with
\bea
G_{a b}^{(0)}(z) & = & \delta(1-z) \,\delta_{ag}\, \delta_{bg} \, , \\
G_{g g}^{(1)}(z) & = & \delta(1-z) \left[
            C_A \, \frac{\pi^2}3  
	  + \beta_0 \ln \left( \frac{\mu_R^2}{\mu_F^2} \right)
          + \sum_{i=0,1/2} {\mathcal G}^{(2l)}_{i} \right] + \dots
\label{real} 
\eea
The dots in Eq.(\ref{real}) represent the contribution from the real
emission that  we have not written as well as the other NLO factors
$G_{g q}^{(1)}(z),\, G_{q q}^{(1)}(z), \, G_{q \bar{q}}^{(1)}(z)$.
In Eq.(\ref{real})    $\beta_0 = (11\, C_A - 4\, n_f \,T(R_f) -
n_s\, T(R_s)/6$ 
 with $n_f\: (n_s)$  the  number of active fermion (scalar) flavor in the 
representation $R_f\: (R_s)$.

The function $ {\mathcal G}^{(2l)}_{i}$ can 
be  cast in the following form:\\
\bea
{\mathcal G}^{(2l)}_{i} & = & 
        \lambda_i \left(\frac{A^2}{m_0^2} \right)^{1-2 i}\, T(R_i) 
               \Biggl( C(R_i)\, {\mathcal G}^{(2l, C_R)}_{i} (x_i)+
                    C_A \,{\mathcal G}^{(2l, C_{A})}_{i} (x_i)  \Biggr) \nn \\
& & \times
     \left( \sum_{j=0,1/2}  \lambda_j \left(\frac{A^2}{m_0^2} \right)^{1-2 j}\,
       T(R_j)\, {\mathcal G}^{(1l)}_{j} \right)^{-1} + h.c.
\label{G2}
\eea
with ${\mathcal G}^{(2l, C_R)}_{i}= {\mathcal F}^{(2l)}_{i} $.
The infrared regularized functions ${\mathcal G}^{(2l, C_{A})}_i$, after 
subtraction of the infrared poles, are found to 
be:
\bea
{\mathcal G}^{(2l, C_{A})}_{1/2} (x)& = & 
      \frac{4x}{(x-1)^2} \Biggl[ \, 3
    + \frac{x (1+8x+3x^2) }{(x-1)^3} \, H(0,0,0,x)  
    - \frac{2 (1+x)^2}{(x-1)^2} \, {\mathcal H}_2 (x)  \nn\\
& & \hspace{21.5mm}  + \, \zeta_3 - H(1,0,0,x) \Biggr]  , 
\label{G12CA} \\
{\mathcal G}^{(2l, C_{A})}_{0} (x) & = & 
           \frac{4x}{(x-1)^2} \Biggl[ - \frac32  
         + \frac{x (1-7x)}{(x-1)^3}   H(0, 0, 0,x)
         + \frac{4x}{(x-1)^2} {\cal H}_2 (x) \Biggr] \, ,
\label{G0CA}
\eea
with
\bea
 {\cal H}_2 (x) & = &
 \frac{4}{5} {\zeta_2}^2 + 2 \zeta_3 
       + \frac{3 \zeta_3}{2} \, H(0,x) 
       + 3 \zeta_3 H(1,x) 
       + \zeta_2 H(1,0,x)
       + \frac{1}{4} \left( 1 + 2 \zeta_2 \right) \,H(0,0,x)   \nn\\
& &    
       -2 \, H(1,0,0, x) + H(0,0,-1,0,x) 
       + \frac{1}{4} H(0,0,0,0,x)
       + 2 \, H(1,0,-1,0,x)  \nn \\  
& &    
       - H(1,0,0,0,x) \, .
\eea
The analytic expression of ${\mathcal G}^{(2l)}_{1/2}$ is in agreement with 
that reported in Ref.\cite{HK} based on the results presented in 
Ref.\cite{QCDg2}.
\section{Pseudoscalar Higgs: $A \to \gamma \gamma$ and $gg \to A$ \label{sec4}}
To complete an independent check of the results of Ref.\cite{QCDg2}
in this section we consider  the virtual NLO QCD corrections to the decay 
width of a pseudo-scalar particle $A$ in two photons, 
$\Gamma(A \to \gamma \gamma)$, and to its production cross section via gluon 
fusion, $\hat{\sigma}(gg \to A)$. 

As in Ref.\cite{QCDg2} we assume the interaction of the $A$ particle with
gluons mediated only by the top quark. Because the NLO QCD corrections to 
these two processes are calculated in
Dimensional Regularization,  a prescription for the $\gamma_5$
matrix is needed.  We use the same prescription of Ref.\cite{QCDg2}, i.e.
't~Hooft--Veltman one \cite{HV}, that, as it is well known, breaks manifestly 
Ward Identities. The latters need 
to be restored explicitly, with a finite renormalization. If $Z_{Hf \bar{f}}$
and $Z_{Af \bar{f}}$ are the renormalization constants of the vertex 
$Hf \bar{f}$, scalar Higgs-fermion-antifermion, and $Af \bar{f}$, 
pseudo-scalar Higgs-fermion-antifermion, respectively, the contribution of 
the finite renormalization can be found using \cite{QCDg2,Larin}:
\be
Z_{Af \bar{f}} = Z_{Hf \bar{f}} + 2 C_F \frac{\alpha_S}{\pi} \, .
\ee

\subsection{Decay Width $A \to \gamma \gamma$}

In analogy with Eq.(\ref{width}), we write:
\be
\Gamma(A \to \gamma \gamma) = \frac{G_{\mu} \alpha^2 m_A^3}{128 \sqrt{2}
\pi^3} \left| {\mathcal E} \right| ^2 ~.
\label{widthA}
\ee
Assuming  the strength of the coupling of the
pseudoscalar to the top quark equal to 
$Att = g\, \eta_t\, \mt/(2\, \mw)$, with $\mt$ the top 
mass and $\eta_t$ a numerical coefficient, the  function ${\mathcal E}$ can be 
written as ($Q_t = 2/3$):
\be
{\mathcal E} =    \eta_t\, Q_{t}^2 \, N_{c} \Bigl[ 
{\mathcal E}^{(1l)}_{t} + \frac{\alpha_S}{\pi} C_F      
{\mathcal E}^{(2l)}_{t} + \dots \Bigr] \, .
\ee

The leading order term is 
\be
{\mathcal E}^{(1l)}_{t} = 4\,y_{t} H(0,0,x_{t}) \, ,
\label{A1loop}
\ee
where $y_t$, $x_t$ are given by Eq.(\ref{defx}) with $i=t$.
At the NLO, assuming  an $\overline{\mbox{MS}}$ top mass we have:
\be
{\mathcal E}^{(2l)}_{t}  =  {\mathcal E}_{t}^{(2l,a)}(x_{t}) 
                   + {\mathcal E}_{t}^{(2l,b)}(x_{t}) 
                   \ln \left({m_{t}^2 \over \mu^2} \right) \, ,
\ee
where:
\bea
{\mathcal E}_{t}^{(2l,a)} (x) & = & 
- \frac{4 x}{(x-1)^2} \bigl[ \zeta_3 - 4 H(0, -1, 0,x) + H(0, 1, 0,x) - 5 H(1, 0, 0,x) \bigr] \nn\\
& & 
+ \frac{4 x \left[ 2(1-x)^2 - \zeta_2(1-x^2) \right]}{(x-1)^3 (1 + x)}  H(0,x)
+ \frac{8 x (1 - x^2)}{(x-1)^3 (1 + x)}  H(0, 0,x) \nn\\
& & 
+ \frac{6 x (1 + x)}{(x-1)^3} H(0, 0, 0,x)
- \frac{8 x (1 + x^2)}{(x-1)^3 (1 + x)} {\mathcal H}_1
\, ,
\label{2lQCDa}\\
{\mathcal E}_{t}^{(2l,b)} (x) & = & -\frac{6x}{(x-1)(1+x)} H(0,x)
+ \frac{6x}{(x-1)^2} H(0,0,x) \, .
\label{2lQCDpolea}
\eea
The corresponding expression for an OS  top  mass is given by:
\be
 {\mathcal E}_{t}^{(2l)} =  {\mathcal E}_{t}^{(2l,a)} (x_{t})
                   + \frac43 {\mathcal E}_{t}^{(2l,b)}(x_{t})~. 
\label{2lQCDosA}
\ee
\subsection{Production Cross Section $gg \to A$}
The expressions for the relevant quantities in the $gg \to A $ production
cross section can be easily obtained from those in Section \ref{sec3}, with
the substitutions: $T(R_i) \to 1/2,\: C(R_i) \to C_F,\: {\mathcal F} \to 
{\mathcal E},\: {\mathcal G}\to {\mathcal K}$. In particular, the Born-level
partonic cross section (Eq.(\ref{ggh})) is: 
\be
\sigma^{(0)}  =  
\frac{G_\mu \alpha_s^2 (\mu_R^2)}{128\, \sqrt{2} \, \pi}
\left| \frac12 \eta_t\,{\mathcal K}^{(1l)}_{t}
       \right|^2 \, ,
\label{ggA}
\ee
with ${\mathcal K}^{(1l)}_{t}= {\mathcal E}^{(1l)}_{t}$. The NLO virtual 
contribution to the gluon fusion subprocess (Eq.(\ref{real})) is:
\be
G_{g g}^{(1)}(z)  =  \delta(1-z) \left[
C_A \, \frac{\pi^2}3  +
    \beta_0 \ln \left( \frac{\mu_R^2}{\mu_F^2} \right)
 +  {\mathcal K}^{(2l)}_{t} \right] \, ,
\ee
where $\beta_0 = (11\, C_A - 2\, n_f)/6$, with $n_f$  the  number of active 
flavor, and  
\bea
{\mathcal K}^{(2l)}_{t}  =  
\left(  {\mathcal K}^{(1l)}_{t} \right)^{-1}
 \Biggl( C_F\, {\mathcal K}^{(2l, C_F)}_{t} (x_t)+
      C_A \,{\mathcal K}^{(2l, C_{A})}_{t} (x_t)  \Biggr) 
 + h.c.
\label{K2}
\eea
with ${\mathcal K}^{(2l, C_F)}_{t}= {\mathcal E}^{(2l)}_{t} $ and
\be
{\mathcal K}^{(2l, C_{A})}_{t} (x) = 
\frac{4 x}{(x-1)^2} \bigl[ \zeta_3 - H(1,0,0,x) - 
2 {\mathcal H}_2 (x) \bigr]
+ \frac{12 x^2}{(x-1)^3} H(0,0,0,x) \, .
\label{K12CA}
\ee
Eqs.(\ref{2lQCDosA},\ref{K2}) are  in agreement with the corresponding 
expressions presented  in Ref.\cite{HK}.

\section{Conclusions}

In this paper, we considered the virtual NLO QCD corrections to the processes 
$H \to  \gamma \gamma$ and $gg \to H$. We assumed  the coupling of the 
Higgs boson to the  photons and gluons  to be mediated by  fermionic and
scalar loops. We provided analytic formulas for these corrections that
are valid for arbitrary mass of the fermion or scalar particle
running in the loops and of the Higgs boson. They are
given in a very compact form as a combination of HPLs.

The calculation here presented was done using the Laporta algorithm for the
reduction of the scalar integrals to the MIs and the differential equations
method for the evaluation of the latters. A part of the MIs needed for the 
calculation was already known in the literature. We explicitly give the 
analytic results for the MIs that were not known.

We checked our  results for the decay width of the Higgs boson in two photons 
and for the partonic cross section of the gluon fusion by  performing 
an independent calculation in the  region of small Higgs mass via an
asymptotic expansion in the variable $r \equiv \mh^2/m^2 \ll 1$, with $m$ 
the mass of the fermion or scalar particle, up to the
first 4-5 orders.

We considered also the NLO virtual QCD corrections to $A \to \gamma \gamma$ 
and $gg \to A$  assuming the coupling of the pseudoscalar boson to the
external particles mediated by a fermion. 

We find complete agreement with the results previously known in the
literature concerning the production and decay of a   
(pseudo)scalar Higgs boson mediated by fermionic loops. 
This provides an independent check of the formulas given in 
Refs.\cite{QCDg2,HK} and extends them to the case of a scalar particle 
running  in the loops.

\vspace*{6mm}

\noindent {\large{\bf Acknowledgments}}

\vspace*{2mm}

\noindent 
The authors want to thank M.~Spira and A.~Djouadi for useful communications
regarding Ref.\cite{QCDg2}. 
R.~B. wishes to thank the Department of Physics of the University of 
Florence and INFN Section of Florence for kind hospitality,
and in particular S. Catani for useful discussions during a large part 
of this work. Discussions with G. Rodrigo are gratefully acknowledged.
The work of R.~B. was partially supported by Ministerio de
Educaci\'on y Ciencia (MEC) under grant FPA2004-00996, 
Generalitat Valenciana under grant GV05-015, and MEC-INFN agreement.

\vspace*{6mm}

\noindent {\large{\bf Note added}}

\vspace*{2mm}

\noindent 
After our work was completed a paper on a similar subject has appeared 
on the Web \cite{Anastasiou:2006hc}. We did not
check yet our formulas against theirs.


\begin{appendletterA}
\section*{Appendix A: Results  in the Low and High Higgs Mass Regimes}

In this Appendix we present approximate results 
that are valid  in regions in which the mass of the Higgs boson is either  
much smaller or much larger than that of the particle running in the loop.

\subsection*{A.1 $H \to \gamma\gamma$}

The expansions of Eqs.(\ref{eq:2}-\ref{eq:4}) with respect to 
$r_i=\mh^2/m_i^2$, when $r_i \ll 1$ read:
\bea
{\cal F}_{1}^{(1l)}(r) & = & 
        		         7
        		       + \frac{11}{30} r
        		       + \frac{19}{420} r^2
        		       + \frac{29}{4200} r^3
        		       + \frac{41}{34650} r^4 
			       + {\mathcal O}\left( r^{5} \right) \, , \\
{\cal F}_{1/2}^{(1l)}(r) & = & - \frac{4}{3} 
                               - \frac{7}{90} r
			       - \frac{1}{126} r^2
			       - \frac{13}{12600} r^3
			       - \frac{8}{51975} r^4
			       + {\mathcal O}\left( r^{5} \right) \, , \\
{\cal F}_{0}^{(1l)}(r) & = &   - \frac{1}{3} 
                               - \frac{2}{45} r
			       - \frac{1}{140} r^2
			       - \frac{2}{1575} r^3
                               - \frac{1}{4158} r^4
			       + {\mathcal O}\left( r^{5} \right) \, ,
\eea
while in the opposite case, $r_i \gg 1$ we find:
\bea
{\cal F}_{1}^{(1l)}(r) & = &    2
                          + \left( 12 - 6 \ln^2(-r) \right) \frac{1}{r}
		+ \left( 24 \ln(-r) + 12 \ln^2(-r) \right) \frac{1}{r^2}
			  + {\mathcal O}\left( \frac{1}{r^3} \right) \, , 
\label{th1} \\
{\cal F}_{1/2}^{(1l)}(r) & = & 
                          - \left( 8 - 2 \ln^2(-r) \right) \frac{1}{r}
		- \left( 8 \ln(-r) + 8 \ln^2(-r) \right) \frac{1}{r^2}
			  + {\mathcal O}\left( \frac{1}{r^3} \right) \, , 
\label{th2} \\
{\cal F}_{0}^{(1l)}(r) & = &
                          \frac{4}{r}
			  + 4 \ln^2(-r) \frac{1}{r^2}
			  + {\mathcal O}\left( \frac{1}{r^3} \right) \, .
\label{th3} 
\eea

The expansions of Eqs.(\ref{2lQCDos},\ref{2lscaos}) when $r_i \ll 1$ are
\bea
{\cal F}_{1/2}^{(2l,a)}(r)  & = & 1
			   - \frac{19}{270} r
			   - \frac{104}{14175} r^2
			   - \frac{6313}{15876000} r^3
			   + \frac{5083}{74844000} r^4
			   + {\mathcal O}\left( r^{5} \right) , \\
{\cal F}_{1/2}^{(2l,b)}(r)  & = & 
			   - \frac{7}{60} r
			   - \frac{1}{42} r^2
			   - \frac{13}{2800} r^3
			   - \frac{16}{17325} r^4
			   + {\mathcal O}\left( r^{5} \right) , \\
{\cal F}_{0}^{(2l,a)}(r)  & = &
        		   - \frac{3}{4}
        		   - \frac{29}{216} r
        		   - \frac{4973}{226800} r^2
        		   - \frac{3137}{882000} r^3
        		   - \frac{1180367}{2095632000} r^4
			   + {\mathcal O}\left( r^{5} \right) , \\
{\cal F}_{0}^{(2l,b)}(r)  & = &
        		   - \frac{1}{4}
        		   - \frac{1}{15} r
        		   - \frac{9}{560} r^2
        		   - \frac{2}{525} r^3
        		   - \frac{5}{5544} r^4
			   + {\mathcal O}\left( r^{5} \right) ,\\
\eea
while in the opposite regime we have:
\bea
{\cal F}_{1/2}^{(2l,a)}(r)  & = & - \Bigl[
            36
          + \frac{36}{5} \zeta^2(2)
          - 4 \zeta(3)
          - 4 ( 1 + \zeta(2) + 4 \zeta(3) ) \, \ln( - r)
          - 4 ( 1 - \zeta(2) ) \, \ln^2( - r) \nn\\
& & 
          + \ln^3( - r)
          + \frac{1}{12} \, \ln^4( - r)
	     \Bigr] \frac{1}{r}
       - \Bigl[
            48
          + 8 \zeta(2)
          - \frac{144}{5} \zeta^2(2)
          - 16 \zeta(3) \nn\\
& & 
          + 16 ( 1 + 4  \zeta(3) ) \, \ln(-r)
          + 2 ( 11 - 8 \zeta(2) ) \, \ln^2(-r)
          - 12 \, \ln^3(-r) \nn\\
& & 
          - \frac{1}{3} \, \ln^4(-r)
	     \Bigr] \frac{1}{r^2}
          + {\mathcal O} \left( \frac1{r^{3}} \right) 
\, , 
\label{th4} \\
{\cal F}_{1/2}^{(2l,b)}(r)  & = &  \left[
            12
	  + 6 \, \ln(-r)
	  - 3 \, \ln^2(-r)
	     \right] \frac{1}{r}
       - \left[
            12
	  - 24 \, \ln^2(-r)
	     \right] \frac{1}{r^2}
          + {\mathcal O} \left( \frac1{r^{3}} \right) 
\, , 
\label{th5} \\
{\cal F}_{0}^{(2l,a)}(r) & = &  \left[
            14
	  - 3 \, \ln(-r)
	     \right] \frac{1}{r}
       + \Bigl[
            6
          - \frac{72}{5} \zeta^2(2)
          - 24 \zeta(3)
          - 8 ( 1 - \zeta(2) - 4 \zeta(3) ) \, \ln(-r) \nn\\
& &
          + ( 17 - 8 \zeta(2) ) \, \ln^2(-r)
          - \frac{5}{3} \, \ln^3(-r)
          - \frac{1}{6} \, \ln^4(-r)
	     \Bigr] \frac{1}{r^2}
          + {\mathcal O} \left( \frac1{r^{3}} \right) 
\, , 
\label{th6} \\
{\cal F}_{0}^{(2l,b)}(r) & = & \left[
	    6 \, \ln(-r)
	  - 3 \, \ln^2(-r)
	     \right] \frac{1}{r^2}
          + {\mathcal O} \left( \frac1{r^{3}} \right) 
\, .
\label{th7} 
\eea
The imaginary part of the expressions in Eqs.(\ref{th1}--\ref{th3}) and
(\ref{th4}--\ref{th7}) can be easily recovered via the substitution
\be
\ln{(-r-i \epsilon)} \rightarrow \ln{r} - i \pi \, .
\label{analitcont}
\ee

\subsection*{A.2 $gg \to H$}

The expansions of Eqs.(\ref{G12CA},\ref{G0CA}) when $r_i \ll 1$ are
\bea
{\cal G}^{(2l, C_{A})}_{1/2}(r) & = & 
        		       - \frac{5}{3}
        		       - \frac{29}{1080} r
        		       - \frac{1}{7560} r^2
        		       + \frac{29}{168000} r^3
        		       + \frac{3329}{74844000} r^4
			       + {\mathcal O}\left( r^{5} \right) \, , \\
{\cal G}^{(2l, C_{A})}_{0}(r) & = &  
        		       - \frac{1}{6}
        		       - \frac{1}{135} r
        		       + \frac{1}{15120} r^2
        		       + \frac{29}{189000} r^3
        		       + \frac{1433}{29937600} r^4
			       + {\mathcal O}\left( r^{5} \right) 
\, ,
\eea
while in the opposite regime we have:
\bea
{\cal G}^{(2l, C_{A})}_{1/2}(r) & = & 
       - \Bigl[
            12
          - \frac{32}{5} \zeta^2(2)
          - 12 \zeta(3)
          + 12 \zeta(3) \ln(-r)
          - ( 1 + 2 \zeta(2) ) \ln^2(-r) \nn\\
& &
          - \frac{1}{12} \ln^4(-r)
          \Bigr] \frac{1}{r}
       + \Bigl[
            28
          + 8 \zeta(2)
          - \frac{128}{5} \zeta^2(2)
          - 64 \zeta(3)
          + 8 ( 1 + 6 \zeta(3) ) \ln(-r) \nn\\
& &
          - 2 ( 1 + 4 \zeta(2) ) \ln^2(-r)
          - \frac{4}{3} \ln^3(-r)
          - \frac{1}{3} \ln^4(-r)
          \Bigr] \frac{1}{r^2}
          + {\mathcal O} \left( \frac1{r^{3}} \right) 
\, , 
\label{th8} \\
{\cal G}^{(2l, C_{A})}_{0}(r) & = & 
          \frac{6}{r}
       + \Bigl[
           \frac{64}{5} \zeta^2(2)
          + 32 \zeta(3) 
          - 24 \zeta(3)  \ln(-r)
          + 2 ( 1 + 2 \zeta(2) ) \ln^2(-r) \nn \\
& & 
          + \frac{2}{3} \ln^3(-r)
          + \frac{1}{6} \ln^4(-r)
          \Bigr] \frac{1}{r^2} 
          + {\mathcal O} \left( \frac1{r^{3}} \right) 
\, , 
\label{th9} 
\eea
The imaginary part of the expressions in Eqs.(\ref{th8},\ref{th9})
can be easily recovered via the substitution of Eq.(\ref{analitcont}).

\subsection*{A.3 $A \to \gamma \gamma$ and $gg \to A$}
The expansion of the one-loop result (Eq.(\ref{A1loop})) when 
$r \equiv m_A^2/\mt^2 \ll 1$ is 
\be
{\mathcal E}_t^{(1l)}(r) = -2 -\frac16 r -\frac1{45}r^2 -\frac1{280} r^3 
        - \frac1{1575} r^4 + {\mathcal O}\left( r^{5} \right) \, , 
\ee
while for $r \gg 1$: we have:
\bea
{\mathcal E}_t^{(1l)}(r) & = &  
            2 \ln^2(-r) \frac{1}{r}
          - 8 \ln(-r)\frac{1}{r^2}
          + {\mathcal O} \left( \frac1{r^{3}} \right) \, .
\eea
The expansions of Eqs.(\ref{2lQCDa},\ref{2lQCDpolea}) when 
$r \ll 1$ read
\bea
{\mathcal E}_t^{(2l,a)}(r)  & = & 
        		       - \frac{1}{3} r
        		       - \frac{19}{360} r^2
        		       - \frac{3533}{453600} r^3
        		       - \frac{151}{141120} r^4
			       + {\mathcal O}\left( r^{5} \right) \, , \\
{\mathcal E}_t^{(2l,b)}(r)  & = & 
        		       -  \frac{1}{4} r
        		       - \frac{1}{15} r^2
        		       - \frac{9}{560} r^3
        		       - \frac{2}{525} r^4
			       + {\mathcal O}\left( r^{5} \right) \, , 
\eea
while in the opposite regime we have:
\bea
{\mathcal E}_t^{(2l,a)}(r)  & = & 
       - \Bigl[
            \frac{36}{5} \zeta_2^2
          - 4 \zeta_3
	  + 4 ( 2 - \zeta_2 - 4 \zeta_3 ) \ln(-r)
          - 4 ( 1 - \zeta_2 ) \ln^2(-r)
          + \ln^3(-r) \nn\\
& & 
          + \frac{1}{12} \ln^4(-r)
          \Bigr] \frac{1}{r}
       - \Bigl[
            24
          + 8 \zeta_2
          + 32 \zeta_3
	  + 8 ( 3 - 2 \zeta_2 )\ln(-r)
          - 22 \ln^2(-r) \nn\\
& & 
          - \frac{8}{3} \ln^3(-r)
          \Bigr] \frac{1}{r^2}
          + {\mathcal O} \left( \frac1{r^{3}} \right) 
\, ,  
\label{th10} \\
{\mathcal E}_t^{(2l,b)}(r)  & = & 
        \Bigl[
            6 \ln(-r)
          - 3 \ln^2(-r)
          \Bigr] \frac{1}{r}
       - \Bigl[
            12
          - 24 \ln(-r)
          \Bigr] \frac{1}{r^2}
          + {\mathcal O} \left( \frac1{r^{3}} \right) 
\, .
\label{th11} 
\eea

For Eq.(\ref{K12CA}) in the limit $r \ll 1$, finally, we have:
\be
{\mathcal K}^{(2l, C_{A})}_t (r) = 
        		       - 2
        		       - \frac{1}{24} r
        		       + \frac{29}{60480} r^3
        		       + \frac{53}{378000} r^4
			       + {\mathcal O}\left( r^{5} \right) \, , 
\ee
and for $r \gg 1$:
\bea
{\mathcal K}^{(2l, C_{A})}_t (r) & = & 
         \Bigl[
            \frac{32}{5} \zeta_2^2
          + 12 \zeta_3
          - 12 \zeta_3 \ln(-r)
          +  ( 1 + 2 \zeta_2 ) \ln^2(-r)
          + \frac{1}{12} \ln^4(-r)
          \Bigr] \frac{1}{r} \nn\\
& & 
       + \Bigl[
            28
          + 8 \zeta_2
          + 8 \ln(-r)
          + 2 \ln^2(-r)
          \Bigr] \frac{1}{r^2}
          + {\mathcal O} \left( \frac1{r^{3}} \right) 
\, .
\label{th12} 
\eea

The imaginary part of  the expressions in 
Eqs.(\ref{th10},\ref{th11},\ref{th12})
can be easily recovered via the substitution of Eq.(\ref{analitcont}).

\end{appendletterA}

\begin{appendletterB}
\section*{Appendix B: Master Integrals for the Two-Loop QCD Corrections}

This Appendix is devoted to the analytic expressions of the MIs
involved in the calculation. They are 11, as it is shown in 
Fig.~\ref{fig1app1}.

\begin{figure}
\bc
\epsfig{file=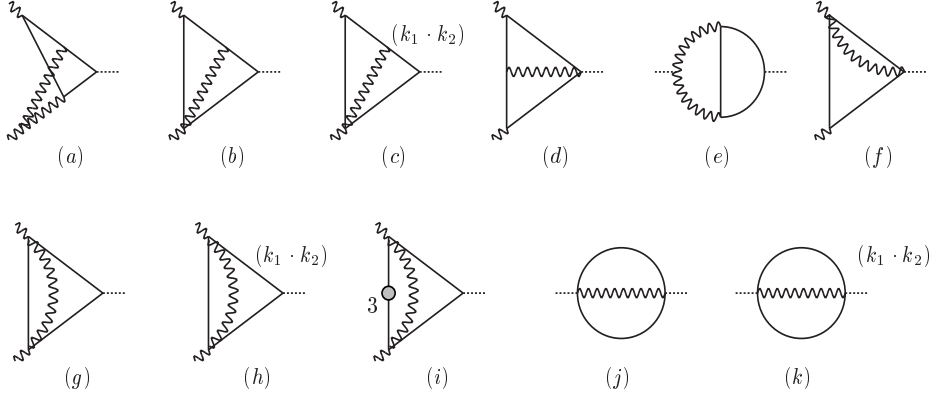,height=5.5cm,width=4.5cm,
        bbllx=220pt,bblly=580pt,bburx=375pt,bbury=763pt}
\caption{\label{fig1app1} \it The Master Integrals necessary for the 
computation of the two-loop
QCD corrections to $gg \to H$ and $H \to \gamma \gamma$.}
\ec
\end{figure}

The diagrams (e) and (g)--(k) can be found in \cite{MIs}. Diagrams (a)--(d)
and (f) are listed below.
The integrals are performed in Euclidean $D$-dimensional space and expanded in 
Laurent series of $\epsilon = (2 - D/2)$. The mass of the particles running in 
the loops is denoted by $m$, while $\mu$ is the unit of mass of the Dimensional
Regularization. The variable $x$ is defined in Eq.~(\ref{defx}).

A common prefactor of
\be
\left( \frac{\mu^{2}}{m^2} \right) ^{2 \epsilon} 
\pi^{2(2-\epsilon)} \, \Gamma^2(1+\epsilon) 
\ee
is understood and not explicitly shown in the formulas below.

All the results of this section can be obtained in an electronic form by 
downloading the source files of this manuscript from http://www.arxiv.org.

\subsection*{B.1 Topology $t=4$ \label{4den}}

\vspace*{5mm}
\be
\hbox{
\begin{picture}(0,0)(0,0)
\SetScale{.35}
  \SetWidth{1.5}
\Photon(-60,60)(-48,50){4}{2}
\Photon(-60,-60)(-48,-50){4}{2}
\DashLine(20,0)(40,0){2}
\Line(-48,50)(20,0) 
\Line(-48,-50)(20,0)  
\Line(-48,-50)(-48,50)
\PhotonArc(12,60)(60,189,278){4}{11}
\end{picture}}  \hspace*{6mm} = 
\sum_{i=-2}^{1} \epsilon^{i} F^{(1)}_{i} + {\mathcal O} \left( 
\epsilon^{2} \right) , 
\ee

\vspace*{5mm}
where:
\bea
F^{(1)}_{-2} & = & \frac{1}{2} \, , \\
F^{(1)}_{-1} & = & \frac{1}{2} \, , \\
F^{(1)}_{0} & = & 
           - \frac{5}{2} - H(0,0,x) 
       - \frac{x}{(x-1)^2} \bigl[
            4 \zeta(3)
          - 2 H(0,0,0,x)
          - 4 H(1,0,0,x) \bigr] \nn\\
& & 
       + \frac{2(1+x)}{(x-1)} H(0,x) , \\
F^{(1)}_{1} & = & 
          - \frac{35}{2} \! 
       - 2 H(0,1,0,x) 
          + \! \frac{1}{(x-1)^2} \Bigl\{
	    \frac{8}{5} \zeta^2(2) x
	    + \! ( 3 x^2 \! - \! 10 x \! + \! 3) \zeta(3)
	    - 2(x^2-1) \zeta(2) \nn\\
& & 
       + \! \bigl[ 12 (x^2-1)\!  + \! (x-1)^2 \zeta(2) - \! 6 x \zeta(3) \bigr] H(0,x) 
       - [
            7
          - 2 x
          - 9 x^2\! 
          + \! 2 x \zeta(2)
          ] H(0,0,x)  \nn\\
& & 
       + \bigl[ 4(x^2-1)- 4 x \zeta(2) \bigr] H(1,0,x) 
       - 12 (x^2-1) H(-1,0,x) 
       - [ 3 x^2 - 8 x + 3 ] \times  \nn\\
& & 
         \times H(0,0,0,x) \! 
       + \! 6 (x-1)^2 H(0,-1,0,x)\!  
       + \! 2 (1 \! + \! x^2) H(1,0,0,x)
           - x \Bigl[
         12 \zeta(3) H(1,x) \nn\\
& &  
       + 12 H(0,0,-1,0,x)
       - 6 H(0,0,0,0,x)
       - 4 H(0,0,1,0,x)
       + 4 H(0,1,0,0,x) \nn\\
& & 
       + 24 H(1,0,-1,0,x)\! 
       - \! 12 H(1,0,0,0,x)\! 
       - \! 8 H(1,0,1,0,x)\! 
       + \! 8 H(1,1,0,0,x) \Bigr] \Bigr\}  .
\eea

\subsection*{B.2 Topology $t=5$ \label{5den}}

\vspace*{8mm}
\be
\hbox{
\begin{picture}(0,0)(0,0)
\SetScale{.35}
  \SetWidth{1.5}
\Photon(-60,60)(-48,50){4}{2}
\Photon(-60,-60)(-48,-50){4}{2}
\DashLine(20,0)(40,0){2}
\Line(-48,50)(20,0) 
\Line(-48,-50)(20,0)  
\Line(-48,-50)(-48,50)
\Photon(-48,0)(20,0){4}{8}
\end{picture}}  \hspace*{6mm}   =  
F^{(2)}_{0} + {\mathcal O} \left( 
\epsilon \right) , 
\ee

\vspace*{8mm}
where:
\bea
m^2 F^{(2)}_{0} & = &  \frac{x}{(x-1)^2} \Bigl\{
          - \frac{12}{5} \zeta^2(2)
       - 4 \zeta(3) [ H(0,x) + 2 H(1,x) ] 
       + 2 H(0,0,0,0,x) \nn\\
& & 
       + 4 H(0,1,0,0,x)
       + 4 H(1,0,0,0,x)
       + 8 H(1,1,0,0,x)
	   \Bigr\} \, .
\eea

\vspace*{8mm}
\be
\hbox{
\begin{picture}(0,0)(0,0)
\SetScale{.35}
  \SetWidth{1.5}
\Photon(-60,60)(-48,50){4}{2}
\Photon(-60,-60)(-48,-50){4}{2}
\DashLine(20,0)(40,0){2}
\Line(-48,50)(20,0) 
\Line(-48,-50)(20,0)  
\Line(-48,-50)(-48,50)
\Photon(-48,-50)(-10,22){4}{10}
\end{picture}} \hspace*{6mm} =  
F^{(3)}_{0} + {\mathcal O} \left( 
\epsilon^{2} \right) , 
\ee

\vspace*{8mm}
where:
\bea
\! \! \! \! \! \! \! \! \! \! \! m^2 F^{(3)}_{0} & = &  \frac{x}{(x-1)^2} \Bigl[
          \zeta^{2}(2)
       + \zeta(2) H(0,0,x) 
       + 2 \zeta(2) H(1,0,x) 
       + H(0,0,1,0,x) \nn\\
\! \! \! \! \! \! \! \! \! \! \! & & 
       + 2 H(0,1,0,0,x) \!  
       + \! 3 H(1,0,0,0,x) \! 
       + \! 2 H(1,0,1,0,x) \! 
       + \! 4 H(1,1,0,0,x) \Bigr]
	   .
\eea

\vspace*{3mm}

For the following MI we have chosen the scalar integral with a scalar
product $k_1 \cdot k_2$ on the numerator, $k_1$ and $k_2$ being the two 
momenta of integration. While the results of the other MIs here given do 
not depend on the routing, $F^{(4)}_i$ do. The five denominators of the 
MI under consideration are: $D_1=[k_1^2+m^2]$, $D_2=k_2^2$, 
$D_3=[(p_1-k_1)^2+m^2]$, $D_4=[(p_1-k_1+k_2)^2+m^2]$, 
$D_5=[(p_2+k_1-k_2)^2+m^2]$.

\vspace*{6mm}

\be
\hbox{
\begin{picture}(0,0)(0,0)
\SetScale{.35}
  \SetWidth{1}
\Photon(-60,60)(-48,50){4}{2}
\Photon(-60,-60)(-48,-50){4}{2}
\DashLine(20,0)(40,0){2}
\Line(-48,50)(20,0) 
\Line(-48,-50)(20,0)  
\Line(-48,-50)(-48,50)
\Photon(-48,-50)(-10,22){4}{10}
\Text(5,4)[cb]{{\footnotesize $(k_1 \cdot k_2)$}}
\end{picture}} \hspace*{10mm} =  
\sum_{i=-2}^{1} \epsilon^{i} F^{(4)}_{i} + {\mathcal O} \left( 
\epsilon^{2} \right) , 
\ee

\vspace*{8mm}
where:
\bea
F^{(4)}_{-2} & = & \frac{1}{2} \, , \\
F^{(4)}_{-1} & = & 2 \, , \\
F^{(4)}_{0} & = & 8 + \frac{1}{2(x-1)^2} \Bigl\{
                           3(x^2-1) \zeta(2)
			-  [ 9(x^2-1) + \zeta(2)(1+x^2)] H(0,x) \nn\\
& & 
			+ ( 4 - x + x^2 ) H(0,0,x) 
        		- 3 x^2 H(0,0,0,x) \Bigr\}
	  + \frac{1+x}{2(x-1)} \bigl[
			  4 H(-1,0,x) \nn\\
& & 
			+ H(1,0,x) \bigr]
	  - \frac{1+x^2}{2(x-1)^2} \bigl[
			  2 H(1,0,0,x)
			+ H(0,1,0,x)  \bigr] \, , \\
F^{(4)}_{1} & = & 29 + \frac{1}{2(x-1)^2} \Bigl\{
        		       \frac{2}{5}(8x^2-1) \zeta^2(2)
			     + 14 (x^2-1) \zeta(2)
        		     - (11 - x + 2x^2) \zeta(3) \nn\\
& &
			     - \bigl[ 35(x^2-1) + (8-3x+x^2) \zeta(2) + (1-10x^2) \zeta(3)
			       \bigr] H(0,x) 
			     + \bigl[ (x^2-1) \zeta(2)  \nn\\
& &
                     + 10(1+x^2) \zeta(3) \bigr] H(1,x) 
			     + \bigl[ 19 + 7 x - 18 x^2 
			             + (4x^2-3) \zeta(2) \bigr] H(0,0,x) \nn\\
& & 
 - \bigl[  3(x^2\! -\! 1) -4(1\! +\! x^2) \zeta(2) \bigr] H(1,0,x) 
			     + \! (1 \! + \! x^2) \Bigl[
  \zeta(2) ( 2 H(0,-1,x) \! - \! H(0,1,x) ) \nn\\
& & 
				+ H(0,-1,0,0,x) 
				+ 2 H(0,-1,1,0,x)  
				+ 2 H(0,1,-1,0,x)
				- 2 H(0,1,1,0,x)   \nn\\
& & 
				+ 12 H(1,0,-1,0,x)  
				- 5 H(1,0,0,0,x) 
				- 2 H(1,0,1,0,x) 
				+ 4 H(1,1,0,0,x)  \Bigr]\nn\\
& & 
			     + \! (2 x^2\! - \! x\! + \! 8) H(0,0,0,x)
			     - 2 ( 10 \! - \! 5x + 7x^2) H(0,-1,0,x) 
			     + 2 (1 \! - \! x + 3x^2) \times \nn\\
& &
			       \times H(0,1,0,x) 
			     - (13 - 10 x + 9 x^2 ) H(1,0,0,x)
			     + 2 (9x^2-2) H(0,0,-1,0,x)  \nn\\
& &
			     - 9 x^2 H(0,0,0,0,x) 
			     - (1+5x^2) H(0,0,1,0,x) \Bigr\}
		     - \frac{1+x}{2(x-1)} \bigl[
			       6 \zeta(2) H(-1,x)\nn\\
& &
			     - 34 H(-1,0,x) 
			     + 8 H(-1,-1,0,x)
			     - 3 H(-1,0,0,x)  
			     + 2 H(-1,1,0,x)\nn\\
& & 
			     + 2 H(1,-1,0,x)
			     - 2 H(1,1,0,x)
			     - 3 H(-1,0,0,0,x)
			     - 7 H(0,1,0,0,x) \bigr]
	  \, .
\eea

\subsection*{B.3 Topology $t=6$ \label{6den}}

\vspace*{8mm}
\be
\hbox{
\begin{picture}(0,0)(0,0)
\SetScale{.35}
  \SetWidth{1}
\Photon(-60,60)(-48,50){4}{2}
\Photon(-60,-60)(-48,-50){4}{2}
\Line(-48,50)(20,0) 
\Line(-10,-22)(-48,50) 
\Line(20,0)(-10,-22)  
\DashLine(20,0)(40,0){2}
\Photon(-48,-50)(-10,-22){4}{6}
\Photon(-48,-50)(-10,22){4}{10}
\end{picture}} \hspace*{10mm}  =  
\sum_{i=-1}^{0} \epsilon^{i} F^{(5)}_{i} + {\mathcal O} \left( 
\epsilon^{2} \right) , 
\ee

\vspace*{8mm}
where:
\bea
F^{(5)}_{-1} & = & \frac{4 x^2}{m^4(x-1)^4} \bigl[ - 3 \zeta(3) - \zeta(2) H(0,x)
       - 2 H(0,-1,0,x) + H(0,0,0,x) \bigr] \, , \\
F^{(5)}_{0} & = & \frac{4 x^2}{m^4(x-1)^4} \Bigl[ 
          - \frac{16}{5} \zeta^2(2)
       - 11 \zeta(3) H(0,x)
       - 12 \zeta(3) H(1,x)
       - 5 \zeta(2) H(0,0,x) \nn\\
& & 
       + 4 \zeta(2) H(0,-1,x)
       - 2 \zeta(2) H(0,1,x)
       - 4 \zeta(2) H(1,0,x)
       + 3 H(0,0,0,0,x) \nn\\
& & 
       + 2 H(0,0,1,0,x)
       + 2 H(0,1,0,0,x)
       + 4 H(1,0,0,0,x)
       + 16 H(0,-1,-1,0,x) \nn\\
& & 
       - 10 H(0,-1,0,0,x)
       - 4 H(0,-1,1,0,x)
       - 4 H(0,1,-1,0,x)
       - 8 H(1,0,-1,0,x) \nn\\
& & 
       - 14 H(0,0,-1,0,x) \Bigr]
	  \, .
\eea
\end{appendletterB}

%

\end{document}